\documentclass[prb,twocolumn,showpacs,preprintnumbers]{revtex4-1}
\usepackage{graphicx}
\usepackage{amsmath}
\usepackage{amssymb}
\usepackage{times}
\usepackage{color}
\usepackage{ulem}
\usepackage{physics}
\usepackage{cleveref}
\usepackage{xr}
\usepackage{multirow}
\usepackage{comment}

\newcommand{\nb}{\bar{n}}

\def\d{\delta}

\def\S{\Sigma}
\def\s{\sigma}

\def\w{\omega}
\def\ua{\uparrow}
\def\da{\downarrow}

\def\cc{\hat{c}}

\def\nn{\hat{n}}

\begin{document}

\title{Doped Mott insulator on Penrose tiling}

\author{Shiro Sakai$^1$ and Nayuta Takemori$^{2}$}
\affiliation{
$^1$Center for Emergent Matter Science, RIKEN, Wako, Saitama 351-0198, Japan\\
$^2$Center for Quantum Information and Quantum Biology, Osaka University, Toyonaka, 560-0043, Japan
}
\date{\today}
\begin{abstract}
We study the effect of carrier doping to the Mott insulator on the Penrose tiling, aiming at clarifying the interplay between quasiperiodicity and strong electron correlations. We numerically solve the Hubbard model on the Penrose-tiling structure within a real-space dynamical mean-field theory, which can deal with a singular self-energy necessary to describe the Mott insulator and spatial inhomogeneity.
We find that the strong correlation effect produces a charge distribution unreachable by a static mean-field approximation. 
In a small doping region, the spectrum shows a site-dependent gap just above the Fermi energy, which is generated by a singularly large self-energy emergent from the Mott physics and regarded as a real-space counterpart of the momentum-dependent pseudogap observed in a square-lattice Hubbard model.

\end{abstract}
\maketitle

{\it Introduction ---}
Carrier doping to the Mott insulator drastically changes the electronic structure \cite{imada98}. A typical example is high-temperature superconducting cuprates, which show $d$-wave superconductivity, pseudogap, strange metal, charge-ordering, and Fermi-liquid phases, depending on the doping concentration and temperature \cite{keimer15}.
These drastic changes are considered to be a consequence of severe competition between electrons' kinetic and interaction energies.
Such a strong correlation effect crucially depends on the underlying crystal structure, as exemplified by another prototypical system of organic conductors \cite{kanoda06}.
Given the versatility of the strong correlation effect in periodic crystals, we may expect further diverse phenomena and exotic phases in quasiperiodic systems, where the electronic states are orderly but inhomogeneous.

Quasiperiodic structure manifests itself in quasicrystals \cite{mackay82,shechtman84, levine84}, some of which contain transition-metal or rare-earth elements. 
In fact, a quantum critical behavior was found in an Au-Al-Yb quasicrystal \cite{deguchi12}, suggesting an important role played by electron correlations.  
Recently, superconductivity \cite{kamiya18} and ferromagnetism \cite{tamura21} have also been discovered in other quasicrystals as a realization of electronic long-range order on quasiperiodic lattices.

Though these experimental discoveries have stimulated many theoretical works, the interplay between the quasiperiodicity and strong correlations remains largely unexplored.
This is partly because the lack of periodicity severely limits applicable theoretical methods. 
An exception would be one dimension, where the effect of a quasiperiodic potential \cite{kohmoto83,ostlund83,jagannathan20} on interacting fermions has been studied since early days \cite{hiramoto90,chaves97,vidal99,schuster02}, and has attracted a resurgent attention in recent years \cite{tezuka10,matsuda14,fan20}, particularly in connection with the ultracold-atom experiment \cite{roati08}.

In two and three dimensions, the geometry of the quasiperiodic lattices comes into play \cite{tsunetsugu86,sutherland87,tokihiro88,arai88}.
In this case, various electronic phases have been studied, based on the Heisenberg-type or Hubbard-type models. 
These include metallic \cite{shaginyan13,andrade15,takemura15,shinzaki16,otsuki16,sakai21,sakai22}, magnetic \cite{lifshitz98,wessel03,vedmedenko04,jagannathan04,wessel05,jagannathan07,szallas09,jagannathan12,thiem15,hartman16,koga17,koga20,hauck21,watanabe21}, superconducting \cite{fulga16,sakai17,araujo19,sakai19,cao20,zhang20,nagai20,takemori20,nagai21} and excitonic insulating \cite{inayoshi20} phases, where relations between the local site geometry and the electron density or order parameters have been clarified.
While these ordered states are basically captured by a static mean-field approach, the Mott insulating state, involving a singular electron self-energy, cannot be described by it: We need to take account of dynamical correlation effects in a nonperturbative way.

The real-space dynamical mean-field theory (RDMFT) \cite{metzner89,georges96,potthoff99,snoek08,koga08} is a valuable tool in this regard.
The RDMFT takes into account local dynamical fluctuations in a site-dependent way.
In Ref.~\onlinecite{takemori15}, the Mott transition in the half-filled Hubbard model on the Penrose tiling was studied with this method.
It clarified that the Mott transition indeed occurs at a critical strength of the onsite repulsion $U_c\simeq 11t$, where $t$ is the electron hopping integral between neighboring sites: 
Despite the site dependence of a locally-defined renormalization factor and double occupancy, 
the transition occurs simultaneously for all the sites. 
Note that, in the above case, all the sites are kept to be half-filled due to the electron-hole symmetry. 

In this paper, we explore the {\it doping-induced} Mott transition on the Penrose lattice, using the RDMFT.
Because of the aperiodicity, the doped holes distribute in a site-dependent way, leading to a real-space differentiation of the electronic structure.
This may be contrasted to the momentum-space differentiation observed in a square-lattice Hubbard model and intensively discussed in the literature as a key to the superconducting mechanism in cuprates.
We find that the strong correlation effect totally changes the population tendency of the doped holes  to the local geometries, leading to a charge distribution essentially different from that of weak couplings. 
In the electron density of states, we find a behavior similar to that of the pseudogap in the square-lattice Hubbard model while in our case the size and position of the gap depend on the real-space coordinate.

\begin{figure}[tb]
\center{
\includegraphics[width=0.48\textwidth]{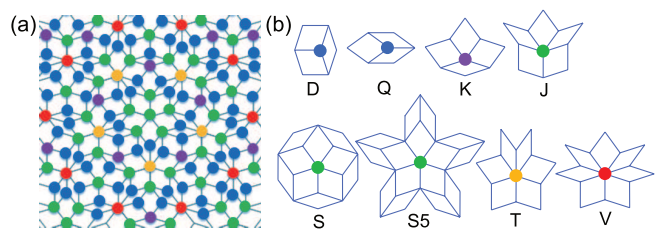}}
\caption{ (a) Penrose tiling. Vertices with a different coordination number $Z_i$ are colored differently: $Z_i=3$ (blue), 4 (purple), 5 (green), 6 (yellow), and 7 (red). (b) Local geometries of the vertices. Notations are after Refs.~\onlinecite{bruijn81-1,bruijn81-2}.}
\label{fig:penrose}
\end{figure}

{\it Model ---}
The Penrose tiling \cite{penrose74} is a prototypical structure of a two-dimensional quasicrystal, covering a plane with only two types of rhombuses [Fig.~\ref{fig:penrose}(a)].
We regard each vertex of a rhombus as a site. 
Each site is in a different environment: In Fig.~\ref{fig:penrose}(a), we classify the sites, according to the coordination number $Z_i$.
The Hubbard Hamiltonian on this lattice reads 
\begin{align}
\hat{H}&=-t \sum_{\langle ij\rangle\s} (\cc_{i\s}^\dagger \cc_{j\s}^{\phantom {\dagger}}+{\it h.c.}) - \mu\sum_{i\s}\nn_{i\s}  
+U\sum_{i}\nn_{i\ua} \nn_{i\da}, 
\label{eq:hubbard}
\end{align}
where $\hat{c}_{i\s}$ ($\cc_{i\s}^\dagger$) annihilates (creates) an electron of spin $\s (=\ua, \da)$ at site $i$ and $\nn_{i\s}\equiv \cc_{i\s}^\dagger \cc_{i\s}$.
The electron hopping $t=1$ is defined between the neighboring two sites (denoted by $\langle ij\rangle$) connected by the edge of the rhombuses.
$U$ represents the strength of the onsite Coulomb repulsion.
The chemical potential $\mu$ is determined self-consistently to fix the average electron density, $\nb\equiv \frac{1}{N}\sum_{i} n_{i}$ with $n_{i}\equiv\sum_{\s}\langle \nn_{i\s}\rangle$ and the system size $N$, at a given value. 
We consider a finite cluster possessing the $C_{5v}$ symmetry, which allows us to deal with a relatively large system of $N=4181$, among which 444 sites are symmetrically inequivalent to each other.
To avoid a peculiarity at the boundary, we use only inner sites satisfying $r<27$ in the plots, where $r$ is the distance from the center of the cluster in a unit of the edge length of the rhombuses.
Note that the Penrose lattice is bipartite: Hence, $n_i\equiv 1$ at half filling ($\nb=1$) and hole and electron dopings are equivalent.

{\it Method ---}
Within the RDMFT, the model (\ref{eq:hubbard}) is mapped onto the Anderson impurity problem at each site, which amounts to 444 different impurity problems in our case.
We then solve each impurity problem, to obtain a site-dependent self-energy $\S_{i}(i\w_n)$ with the Matsubara frequency $\w_n=(2n+1)\pi T$.
Neglecting nonlocal components of the self-energy, we calculate the site-dependent dynamical mean field
\begin{align}
  g_{0,i}^{-1}(i\w_n) =  [\hat{G}(i\w_n)]_{ii}^{-1} +\S_i(i\w_n) \label{eq:g0}
\end{align}
with 
\begin{align}
   [\hat{G}(i\w_n)^{-1}]_{ij} =  \{i\w_n+\mu-\S_i(i\w_n)\}\d_{ij}+t\d_{\langle ij\rangle}.
\end{align}
Here, $\d_{\langle ij\rangle}$ is 1 only when $i$ and $j$ are connected by an edge of a rhombus and 0 otherwise.
Equation (\ref{eq:g0}) defines the Anderson impurity problem in the next step of the self-consistent loop.
To solve the Anderson impurity problem, we use the exact diagonalization (ED) method \cite{caffarel94}, where we represent Eq.~(\ref{eq:g0}) with six bath sites.
Here, we concentrate on a paramagnetic solution at $T=0$. 
The ED method is advantageous to the present study in light of the efficiency in solving each impurity problem, controllability of small doping values, and capability of calculating real-frequency properties.
For real-frequency quantities, we replace $i\w_n$ with $\w+i\eta$, where $\eta=0.01t$ is the energy-smearing factor.

For comparison, we have also implemented calculations with the Hartree-Fock approximation (HFA), using the kernel polynomial method \cite{weisse06}. In this method, the interaction effect is approximated by a site-dependent static potential $Un_i/2$,  so that the self-energy singularity, essential to the Mott physics, is not taken into account.


\begin{figure}[tb]
\center{
\includegraphics[width=0.48\textwidth]{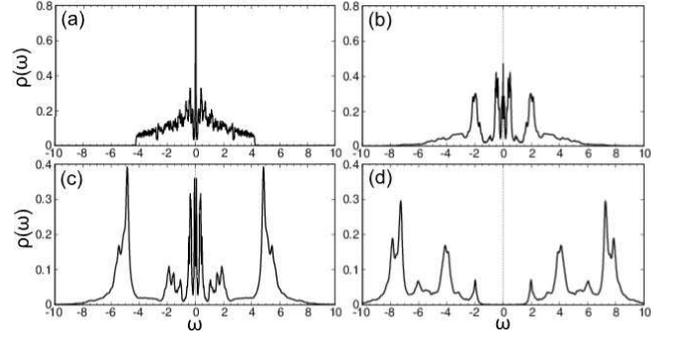}}
\caption{Site-averaged DOS at half filling for (a) $U=0$, (b) $U=4$, (c) $U=8$, and (d) $U=12$, calculated with the RDMFT.} \label{fig:dos}
\end{figure}

\begin{figure}[tb]
\center{
\includegraphics[width=0.48\textwidth]{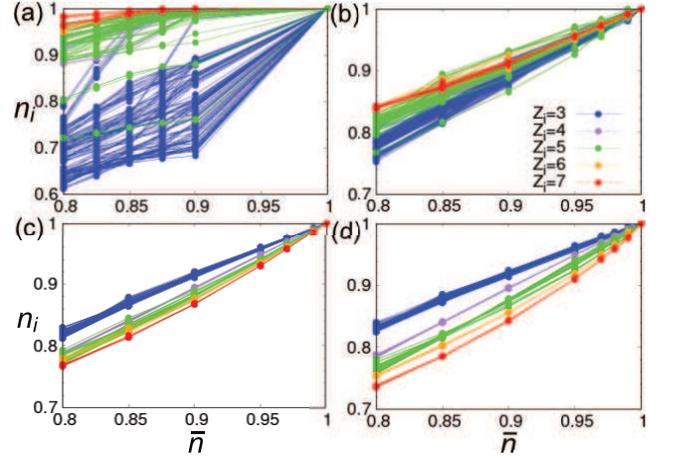}}
\caption{RDMFT results of $n_i$ plotted against the average electron density $\nb$ for (a) $U=0$, (b) $U=4$, (c) $U=8$, and (d) $U=12$.} \label{fig:ni-n}
\end{figure}

\begin{figure}[tb]
\center{
\includegraphics[width=0.48\textwidth]{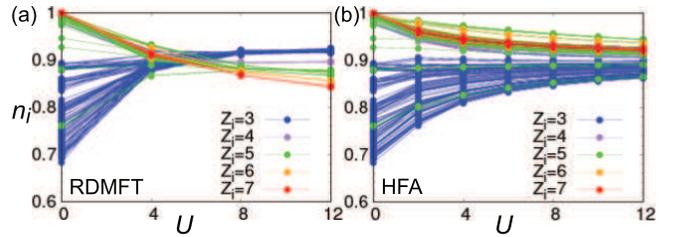}}
\caption{Electron density $n_i$ at each site plotted against $U$, obtained with (a) the RDMFT and (b) the Hartree-Fock approximation (HFA) for $\nb=0.90$, respectively.} \label{fig:ni-U}
\end{figure}

\begin{figure}[tb]
\center{
\includegraphics[width=0.48\textwidth]{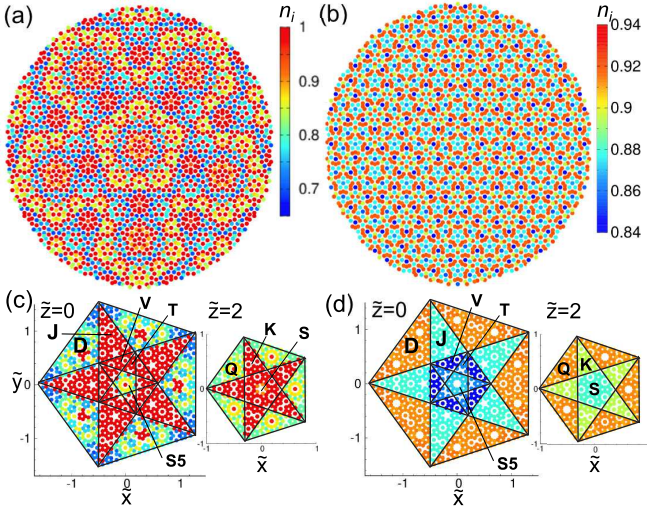}}
\caption{(a), (b) Real-space maps of $n_i$ on the Penrose lattice, calculated with the RDMFT at $\nb=0.90$ for $U=0$ and $U=12$, respectively. (c), (d) Corresponding perpendicular-space maps.
$\tilde{x}$, $\tilde{y}$, and $\tilde{z}$ denote the orthogonal axes of the perpendicular space.} \label{fig:rmap}
\end{figure}

{\it Results ---} We first demonstrate the Mott transition at half filling ($\nb=1$). 
Figure \ref{fig:dos} shows the site-averaged density of states (DOS), $\rho(\w)$, for $U=0$, $4$, $8$, and $12$.
At $U=0$, the "bandwidth" is about $8.5t$, and we can see a $\d$-functional peak of the confined states (discussed below) at $\w=0$.
As $U$ increases, the Hubbard bands develop around $\pm U/2$ and a low-energy spectrum
 loses its weight.
Eventually, at $U=12$, the Mott gap opens.
The critical interaction strength $U_c$ is therefore between $8$ and $12$, in consistency with the result ($U_c\simeq11$) in Ref.~\onlinecite{takemori15}.

Figure \ref{fig:ni-n} plots $n_i$ against $\nb$ for various values of $U$, where we use different colors for different $Z_i$ (same as the color code used in Fig.~\ref{fig:penrose}).
At $U=0$, it is known that the single-electron states are substantially degenerate at half filling. These are localized zero-energy eigenstates of the tight-binding Hamiltonian and called confined states \cite{kohmoto86PRL,arai88}. 
They constitute a $\delta$-functional peak at $\w=0$ in the DOS [Fig.~\ref{fig:dos}(a)] and its weight is calculated to be about $9.8\%$ \cite{kohmoto86PRL,arai88}.
Hence, in Fig.~\ref{fig:ni-n}(a), we consider only $\nb\leq0.9$ and $\nb=1$.
We see that the holes are doped mainly at the $Z_i=3$ sites and some of the $Z_i=5$ sites which possess a local five-fold rotational symmetry.
This is because the confined states have the amplitude only at these sites \cite{arai88}.
At larger doping, the sites with a smaller $Z_i$ tend to be doped more.

The confined states are broken for a finite $U$, where we can study a smaller doping region.
In this region, we can see that all the sites are more or less doped [Figs.~\ref{fig:ni-n} (b)-(d)].
At $U=4$, the sites with a {\it smaller} $Z_i$ tend to be doped more, similarly to the tendency for $U=0$ and $\nb\leq0.9$.
This tendency may be intuitively understood as a narrower level distribution of the sites with a smaller $Z_i$ due to fewer hoppings to neighboring sites.

For $U\geq 8$, we find an opposite tendency: The sites with a {\it larger} $Z_i$ are more doped.
This will be to reduce the interaction energy due to the strong Coulomb repulsion: Since the larger-$Z_i$ sites have more chances to have a transfer of electrons from neighboring sites, it will be preferable to reduce the population of such sites.

The change from weak to strong couplings is clearer in Fig.~\ref{fig:ni-U}(a), which plots $n_i$ against $U$ for $\nb=0.90$.
We find that the population tendency against $Z_i$ is reversed around $U\sim6$.
For $U\lesssim6$, the spread of $n_i$ decreases with $U$.
This is because $U$ prefers a uniform distribution to reduce the onsite potential energy: Suppose two neighboring sites have the populations $n+\d$ and $n-\d$, the interaction energy $\frac{U}{4}\left[(n+\d)^2+(n-\d)^2 \right]$ is minimized at $\d=0$.
As Fig.~\ref{fig:ni-U}(b) shows, this suppression of the spread is captured by the HFA, too\cite{sakai21,sakai22}.
In contrast, for $U\gtrsim6$, the spread of $n_i$ {\it increases} with $U$ in the RDMFT results [Fig.~\ref{fig:ni-U}(a)], where the tendency against $Z_i$ is reversed.
This is a nontrivial strong correlation effect, not captured by the HFA [Fig.~\ref{fig:ni-U}(b)].

These results indicate that the charge distribution for $U\gtrsim 6$ essentially differs from that for $U\lesssim 6$.
We plot $n_i$ in the real space in Figs.~\ref{fig:rmap}(a) ($U=0$) and \ref{fig:rmap}(b) ($U=12$) for $\nb=0.90$. 
Besides the difference in the range of $n_i$, the real-space structures look totally different.

To see the connection with the local geometries, it is useful to plot $n_i$ in the perpendicular space \cite{bruijn81-1,bruijn81-2,jagannathan07}, which consists of the dimensions remaining after projecting the five-dimensional hypercubic lattice onto a two-dimensional plane to construct the Penrose tiling. 
Since one dimension (say, $\tilde{z}$ direction) in the perpendicular space gives a degree of freedom to form different local isomorphism classes, we consider only four planes indexed by $\tilde{z}=0$, $1$, $2$, and $3$. 
When the two sublattices of the Penrose lattice are equivalent, we can concentrate only on $\tilde{z}=0$ and $2$ (see Refs.~\onlinecite{bruijn81-1,bruijn81-2} or \onlinecite{sakai22} for more details).
These are plotted for $n_i$ in Figs.~\ref{fig:rmap}(c) and \ref{fig:rmap}(d).
For $U=0$, we can see that holes are doped at D, Q, and S5 sites of Fig.~\ref{fig:penrose}(b).
These are the sites having the finite amplitude in the confined states \cite{arai88}.
On the other hand, for $U=12$, the holes are more doped at V and T sites while D and Q sites are less doped, making the charge distribution pattern qualitatively different from the noninteracting one.
The fact that $n_i$ is well categorized in the perpendicular space means that the local geometry plays a major role in the charge distribution of the doped Mott insulators.

\begin{figure}[tb]
\center{
\includegraphics[width=0.48\textwidth]{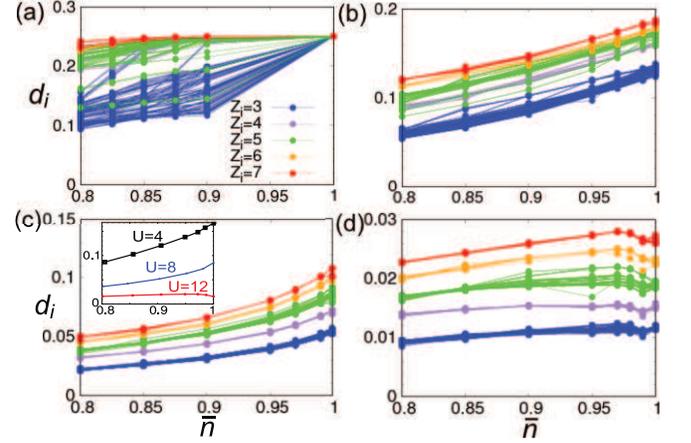}}
\caption{Double occupancy $d_i$ at each site plotted against the average electron density $\nb$, calculated with the RDMFT for (a) $U=0$, (b) $U=4$, (c) $U=8$, and (d) $U=12$. 
Inset to (c) shows the DMFT result for the Bethe lattice with the bandwidth $8$.} \label{fig:di-n}
\end{figure}

Another quantity of interest is the double occupancy, $d_i\equiv\langle \hat{n}_{i\ua}\hat{n}_{i\da}\rangle$, plotted in Fig.~\ref{fig:di-n} against $\nb$ for $U=0$, $4$, $8$ and $12$.
At $U=0$, $d_i$ is equal to $n_i^2/4$, so that it tends to be larger for larger $Z_i$.
This tendency holds for $U=4$, too.
Remarkably, the same holds even for $U=8$ and $12$ despite that $n_i$ acquires an opposite tendency against $Z_i$ (Fig.~\ref{fig:ni-n}).
This indicates a significantly stronger correlation effect at smaller-$Z_i$ sites, where an effective "bandwidth" is small.

Furthermore, the results for $U=8$ and $12$ show qualitatively different behaviors around half filling: While $d_i$ monotonically increases with $\nb$ for $U=8$, it shows a suppression around half filling for $U=12$.
This difference is attributed to the different ground states at half filling: It is a metal for $U=8$ and Mott insulator for $U=12$, as we have seen in Fig.~\ref{fig:dos}.
In fact, a similar behavior is seen in the DMFT result for the Bethe lattice, as is shown in the inset to Fig.~\ref{fig:di-n}(c), where the half-filled electronic state is metallic and Mott-insulating at $U=8$ and $12$, respectively.
Here, the reduction of the double occupancy in the Mott insulator is owing to the disappearance of quasiparticles at the opening of the Mott gap.

Although such quasiparticles, extending on the lattice, are absent on the Penrose lattice, there is a substantial low-energy spectral weight for $0<U<U_c$, which disappears above $U_c$ (Fig.~\ref{fig:dos}).
The electron states corresponding to these low-energy spectra will not be so localized, either:
Aside from the confined states,  the single-electron eigenstates for $U=0$ are considered to be critical \cite{tokihiro88,tsunetsugu91PRB1}.
We therefore think that the reduction of $d_i$ around $\nb=1$ is attributable to the disappearance of such a low-energy electron state.
In fact, in Fig.~\ref{fig:di-n}(d), some sites show a small upturn from $\nb=0.99$ to $\nb=1$, which indicates that the electon states are marginally extended.

\begin{figure}[tb]
\center{
\includegraphics[width=0.48\textwidth]{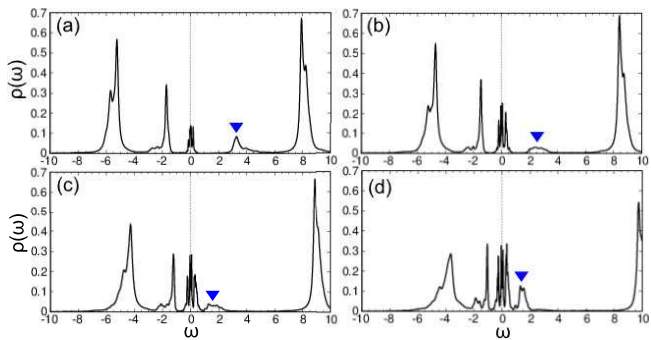}}
\caption{Site-averaged density of states of doped Mott insulators, calculated with the RDMFT for $U=12$ and (a) $\nb=0.99$, (b) $\nb=0.97$, (c) $\nb=0.95$ and (d) $\nb=0.90$.
 Blue triangles denote the ingap state.} \label{fig:dos_dope}
\end{figure}

\begin{figure}[tb]
\center{
\includegraphics[width=0.48\textwidth]{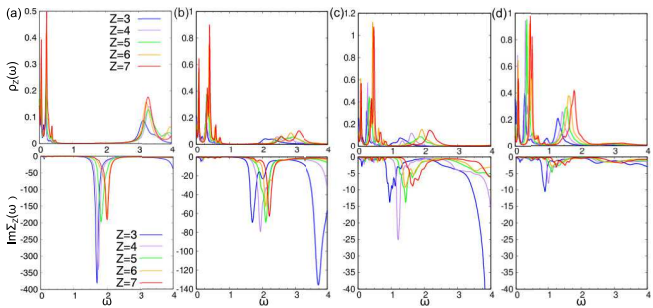}}
\caption{$\rho_Z(\w)$ and Im$\S_Z(\w)$ of doped Mott insulators, calculated with the RDMFT for $U=12$ and (a) $\nb=0.99$, (b) $\nb=0.97$, (c) $\nb=0.95$, and (d) $\nb=0.90$.} \label{fig:dos_dope_z}
\end{figure}

We now turn to the real-frequency properties of the doped Mott  insulators.
Figure \ref{fig:dos_dope} shows the site-averaged DOS at $U=12$.
The corresponding DOS at half filling is shown in Fig.~\ref{fig:dos}(d), which exhibits the Mott gap of about $4t$.
When holes are doped, the chemical potential shifts to the upper edge of the lower Hubbard band  while the Mott gap persists on the unoccupied side [Fig.~\ref{fig:dos_dope}(a)].
As the doping increases, the spectral weight around zero energy increases and the Mott gap shrinks, i.e., the peak denoted by the blue triangle shifts down.

We focus on this low-energy behavior in Fig.~\ref{fig:dos_dope_z}.
The upper panels show the partial density of states, $\rho_Z(\w)$, averaged over the sites with $Z_i=Z$.
We see that a smaller $Z$ tends to give a smaller gap, which is above the Fermi energy but remains finite even down to $\nb=0.90$.
The lower panels show the imaginary part of the local self-energy, Im$\S_Z(\w)$, averaged over the sites with $Z_i=Z$. 
Although this quantity does not directly correspond to the inverse of the $Z$-averaged Green's function, it roughly represents the average behavior of the self-energy at the sites with $Z_i=Z$.
At $\nb=0.99$, we find very strong negative peaks of Im$\S_Z$ slightly below $\w=2$, which is sandwiched by the peaks of $\rho_Z$. 
These peaks of Im$\S_Z$ continuously evolve with doping from the peaks generating the Mott gap at $\nb=1$.
As doping increases, these peaks gradually lose their intensity but are still significant enough to make a gap in $\rho_Z$ at $\nb=0.90$.
While this gap seems to be always above the Fermi energy, the size of the gap and its energy position depends on the real-space coordinate.

{\it Discussion ---} In general, doping the Mott insulator generates a weight just above the Fermi level \cite{eskes91}.
In the single-site DMFT, it is known that this additional weight merges with the quasiparticle peak without making a gap between them and grows as doping increases. Accordingly, the upper Hubbard band gradually loses its weight but does not significantly change its position \cite{georges96,camjayi06}.

Our RDMFT results on the Penrose lattice are different from this conventional DMFT result in that the upper Hubbard band shifts to lower energy (i.e., the Mott gap shrinks) as the doping increases.
This behavior is rather closer to the cluster-DMFT \cite{kotliar01,maier05RMP} results for a square-lattice Hubbard model, where the hole doping causes a substantial reduction of the Mott gap \cite{sakai09PRL,sakai18}, leaving a small gap in a low-energy region.
Because the energy position of this gap strongly depends on momentum \cite{huscroft01,maier02,kyung06PRB1,stanescu06,liebsch09,sakai09PRL,civelli09PRB,lin10,sakai13}, a reduced but finite DOS remains at the Fermi level \cite{civelli05,sordi12PRL,gull13PRL}.
This is called the pseudogap, in analogy with a similar phenomenon observed in cuprate high-temperature superconductors \cite{keimer15}. 
The spectral weight just above the pseudogap is called the ingap state, which {\it shifts} downward with doping from the upper Hubbard band to the low-energy region \cite{sakai18}.
In Fig.~\ref{fig:dos_dope}, the weight denoted by the blue triangle seems to correspond to this ingap state.

As we have noted above, this downward shift of the ingap state does not occur in the single-site DMFT. 
In the cluster DMFT, on the other hand, the self-energy acquires a momentum dependence and this degree of freedom allows the ingap state to shift downward with doping, generating the momentum-dependent pseudogap.
Here, on the Penrose lattice, the self-energy, calculated with the RDMFT, has a dependence on the real-space coordinate.
Although the gap we have seen on the Penrose lattice is located above the Fermi energy, the downward shift of the ingap state with doping, as well as the existence of the gap between the ingap state and the peaks around the zero energy, is common to the pseudogap behavior observed in the cluster DMFT. 

Such a gap just above the Fermi energy would have an interesting consequence on optical conductivity or other spectroscopic and transport properties. In particular, the thermoelectric property is of interest because the large asymmetry of the DOS around the Fermi energy generally gives a large Seebeck coefficient. The role of the self-similarity in these properties is an intriguing future issue.

{\it Summary ---} We have numerically studied the doped Mott insulator on the Penrose tiling through the RDMFT. 
We have shown that the hole distribution drastically changes from weak to strong couplings: At weak couplings, the holes tend to be doped more at the sites with a smaller coordination number while at strong couplings the tendency is reversed.
Since this latter tendency is not captured within the Hartree-Fock approximation, it is a manifestation of a nontrivial interplay between the quasiperiodicity and the strong correlation.
Another interplay is found in the spectrum: The downward shift of the ingap state and the presence of the site-dependent gap above the Fermi energy are characteristic of the doped Mott insulators on the quasiperiodic lattice.

Thus, doped Mott insulators on a quasiperiodic lattice have electronic states which have never been reached before. Their property and possible ordered phases under perturbations pose intriguing issues for future study.


\begin{acknowledgments}
This work was supported by JSPS KAKENHI Grant No. JP16H06345, JP20H05279, 19H05820, and 19H05817.
\end{acknowledgments}

\bibliography{ref}

\end{document}